# Concurrent interactive visualization and handling of molecular structures over the Internet in web browsers

Luciano A. Abriata

Laboratory for Biomolecular Modeling, Swiss Federal Institute of Technology (EPFL)
and Swiss Institute of Bioinformatics, 1015 Lausanne, Switzerland

luciano.abriata@epfl.ch

**Abstract.** This preprint presents a web app (essentially a web page-based program) with which two or more users ("peers") can view and handle 3D molecular structures in a concurrent, interactive way through their web browsers. This means they can share orientation and zoom level, commands and other operations in almost real time over the Internet through standard web pages. This web app, open source and built with the open source components JSmol for molecular visualization and Peer.js for WebRTC connection, provides a practical tool for online collaboration and teaching at a distance. More broadly, it illustrates the strong integrability of technologies for client-side web programming, and paves the way for similar apps for concurrent work in other disciplines. Web app is available at: http://lucianoabriata.altervista.org/jsinscience/concurrent-jsmol/concurrent-jsmol-visualization.html

Concurrent visualization of molecular structures over the Internet is essential for efficient collaboration and teaching over distant locations in disciplines like chemistry and structural biology. A possible way to achieve concurrent visualization is through screen-sharing programs; however, this has a number of disadvantages. First, large amounts of video have to be streamed over the Internet, potentially resulting in slow experience and/or decreased graphics quality for the receiving peer. Second, confidential information could be captured as the video streams through Internet. Third, and probably most important, the receiver has no control over the visualization. All these shortcomings are alleviated by the prototypical web app (program embedded in a web page[1,2]) introduced here, which uses simple methods and existing commodity web software to provide a unique experience for concurrent, interactive molecular visualization and handling over the Internet. As provided, the web app is based on JSmol[3] for molecular visualization and Peer.js[4] for simplified web socket-based connectivity, but other libraries could be used too. (For the user this is anyway irrelevant because no software needs to be installed and the web app works in any standard web browser.)

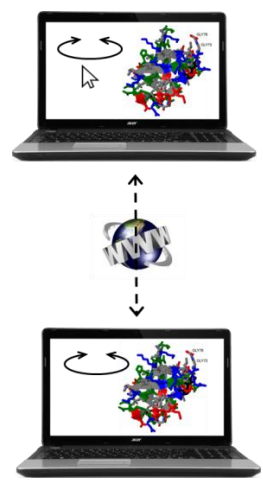

**Figure 1.** Scheme depicting how this web app allows a user to work on a molecular representation and send states and commands seamlessly to the other users, yielding a smooth, computationally inexpensive collaborative experience where the receiver user can also effectively intervene.

## Brief Description for Users

Essentially, this web app provides direct browser-to-browser connectivity to transmit JSmol states (like rotations, zoom levels, etc.) and commands between users. This way, when a user rotates or zooms the structure on his/her screen, this induces the same rotations and zooms on the screens of other users, with subsecond latency. Likewise, a user can run a JSmol command on his/her visualization and apply the same command on all the other users' visualizations. Since the software system transfers only small pieces of text-like information about states and commands, but no graphics, the experience is very fluid even over average Internet connections. Also, the molecular files loaded by each user are never sent over Internet, protecting privacy.

To use the web app, each user (or "peer") must direct his or her browser to http://lucianoabriata.altervista.org/jsinscience/concurrent-jsmol/concurrent-jsmol-visualization.html. When each user's page loads, he or she gets assigned a random "PeerJS ID". At least one user ID must be communicated to the other users (by e-mail or any other medium). When the other users receive the ID of this "master user", they use it to connect to him/her. Users can then select whether they transmit rotations, states and JSmol commands to the master user as they apply them to their visualizations, and whether they apply to their visualizations the rotations, states and commands received from the master user. Independently, any other pair of users can additionally connect to each other to send and receive states and commands directly between them.

Within the web app, users can also chat and send files to each other. And although not implemented in this web app, the communication technology employed also allows the incorporation of video and audio conferencing.

**Brief technicalities about the implementation**

This web app is self-contained in a single HTML file. It uses web sockets (here through the Peer.js library, but others could be used) and a plugin-less molecular viewer for browsers (here JSmol, but also other scriptable viewers could be used). Through HTML and JavaScript code, the web app controls when it sends rotations and commands applied by one peer to the other peers, and whether the rotations and commands received from another user are applied.

The current implementation runs only over http, but web socket libraries that handle https could be used to allow https access (which would enable JSmol to directly load structures directly from the Protein Data Bank, instead of the user having to download it first as in the current version of the web app).

A last important point is that the web app, as provided in its single HTML file, can be directly copied to any other webserver retaining functionality (a copy of JSmol and proper reference to it must be given).

**Perspectives**

Being this web app only a prototype to show feasibility, there are multiple modifications and additions worth exploring. For example, it could be useful to transmit not only commands typed for the JSmol applet but also any other command applied directly from the JSmol interface or console. Also, additional commands prebuilt into buttons could be implemented, depending on the intended uses (small molecules or biological molecules or materials, for research or didactic, etc.)

This example web app shall promote the development of interfaces for concurrent work in other disciplines. Not only for molecular sciences but for virtually any discipline where transmitting small pieces of data to transform views on other users' devices can help to better communicate ideas and concepts at a distance.